\documentstyle[preprint,aps,10pt]{revtex}
\tightenlines
\begin{document}
\draft
\hfill\vbox{\baselineskip14pt
            \hbox{February 2001}}
\baselineskip20pt
\vskip 0.2cm 
\begin{center}
{\Large\bf SET based experiments for HTSC materials II}
\end{center} 
\vskip 0.2cm 
\begin{center}
\large Sher~Alam$^{1}$,~M.~O.~Rahman$^{2}$,~H.~Oyanagi$^{2}$,
and T.~Yanagisawa$^{1}$
\end{center}
\begin{center}
$^{1}${\it Photonics, AIST, Tsukuba, Ibaraki 305, Japan}\\
$^{2}${\it  GUAS \& Photon Factory, KEK, Tsukuba, Ibaraki 305, Japan}\\
$^{3}${\it Quantum Nanoelectronics, AIST, Tsukuba, Ibaraki 305, Japan}

\end{center}
\vskip 0.2cm 
\begin{center} 
\large Abstract
\end{center}
\begin{center}
\begin{minipage}{14cm}
\baselineskip=18pt
\noindent
 The cuprates seem to exhibit statistics, dimensionality
 and phase transitions in novel ways. The nature of excitations
 [i.e. quasiparticle or collective], spin-charge separation, 
 stripes [static and dynamics], inhomogeneities, psuedogap, effect of 
 impurity dopings [e.g. Zn, Ni] and any other phenomenon in these materials
 must be consistently understood. In this note we further discuss
 our original suggestion of using Single Electron Tunneling Transistor 
 [SET] based experiments to understand the role of charge dynamics in 
 these systems. Assuming that SET operates as an efficient charge detection 
 system we can expect to understand the underlying physics of charge 
 transport and charge fluctuations in these materials for a range of 
 doping. Experiments such as these can be classed in a general
 sense as mesoscopic and nano characterization of cuprates
 and related materials. In principle such experiments can
 show if electron is fractionalized in cuprates as indicated
 by ARPES data. In contrast to flux trapping experiments SET
 based experiments are more direct in providing evidence
 about spin-charge separation. In addition a detailed picture
 of nano charge dynamics in cuprates may be obtained.

\end{minipage}
\end{center}
\vfill
\baselineskip=20pt
\normalsize
\newpage
\setcounter{page}{2}
\section{Introduction}
	Keeping the properties of the cuprates and related 
materials in mind [see below], in this short note we further 
discuss our original suggestion to the possibility of  Single Electron 
Tunneling Transistor [SET] based experiments which can probe the 
charge dynamics in HTSC cuprates \cite{alam01-2}, in particular 
the detection of charge-rich and charge-poor [i.e. stripes] and 
the important question of detection of fractional charge $q_{f}~e$ 
carried by Luttinger excitation [{\em The Luttinger excitation is 
fractionalized and the elementary excitation carries the fractional
charge $q_{f}~e$ instead of the quantum of charge $e$ of the electron}].
The HTSC materials are doped Mott insulators, in other words
the parent compounds of HTSC material are unusual insulators.
Superconductivity occurs when they are appropriately doped away
from stoichemistry.

\subsection{Motivation}
	We now outline the theoretical rationale behind our
current proposal, in order to aid in the understanding and
reasons pertaining to our suggestion.
	In a previous work one of us \cite{alam98} has advanced 
the conjecture that one should attempt to model the phenomena of
antiferromagnetism and superconductivity by using quantum
symmetry group. Following this conjecture to model the phenomenona of
antiferromagnetism and superconductivity by quantum symmetry
groups, three toy models were proposed \cite{alam99-1}, namely,
one based on ${\rm SO_{q}(3)}$ the other two constructed with
the ${\rm SO_{q}(4)}$ and ${\rm SO_{q}(5)}$ quantum groups. 
Possible motivations and rationale for these choices  
were outlined \cite{alam99-1}. In \cite{alam99-2} a model to 
describe quantum liquids in transition from 1d to 2d dimensional 
crossover using quantum groups was outlined. In \cite{alam00-1} 
the classical group ${\rm SO(7)}$ was proposed as a toy model to 
understand the connections between the competing phases and the 
phenomenon of psuedo-gap in High Temperature Superconducting Materials
[HTSC]. Then we proposed in \cite{alam00-2} an idea to
construct a theory based on patching critical points so as to
simulate the behavior of systems such as cuprates.
To illustrate our idea we considered an example 
discussed by Frahm et al., \cite{fra98}. The model
deals with antiferromagnetic spin-1 chain doped with
spin-1/2 carriers. In \cite{alam00-3} the connection between
Quantum Groups and 1-dimensional [1-d] structures such as
stripes was outlined. The main point of \cite{alam00-3}
is to emphasize that {\em 1-d structures play an important
role in determining the physical behaviour [such as the 
phases and types of phases these materials are capable of
exhibiting] of cuprates} and related materials.
In order to validate our quantum group conjecture
for the cuprates, we have considered the connection between 
quantum groups and strings \cite{alam01-1,kak91}. Indeed
the 1-d structure can be regarded as one of the basic building 
blocks of a new formulation of correlated condensed matter physics.
In fact we propose this as a definition of what we mean by
a strongly correlated system. Already known example is
when Landau quasiparticles give away to Luttinger liquid
in 1-d. A possible definition of [strongly] correlated electron 
system is as follows: We define a [strongly] correlated electron 
system as one where electron-electron interactions are {\em string-like}:
Examples: BCS, and Stripes.
\subsection{Some properties of cuprates}
	The cuprates seem to exhibit statistics, dimensionality
and phase transitions in novel ways. We summarize the following 
interesting features that appear to arise in these materials,
and motivates us to suggest SET based experiments to understand
charge dynamics in these materials:
\begin{itemize}
\item{}Charge-spin separation:-Indications for
electron fractionalization from Angle-Resolved 
Photoemission Spectroscaopy [ARPES] have
been reported \cite{org00}. We note that the Fermi 
liquid is characterized by sharp fermionic
quasiparticle excitations and has a discontinuity in the
electron momentum distribution function. In contrast
the Luttinger liquid is characterized by charge $e$
spin $0$ bosons and spin $1/2$ charge $0$ and the
fermion is a composite of these [i.e. fractionalization].
It is well-known that transport properties
are defined via correlation functions.
The correlation functions of a Luttinger liquid
have a power law decays with exponents that
depend on the interaction parameters, see below, section II. 
Consequently the transport properties of a Luttinger liquid
are very different from that of a Fermi liquid.
Photoemission experiments on Mott insulating oxides
seems to indicate the spinon and holon excitations
of a charge Luttinger liquid. However the experimental
signatures of Luttinger liquid are not totally
convincing. To this end we propose SET based experiments
to determine the Luttinger liquid behaviour of
the cuprates, see below.  
\item{}d-wave symmetery:-Experimental evidence
for predominantly d-wave pairing symmetry in both hole- 
and electron-doped high T$_{c}$ cuprate superconductors
has been reported by C.C. Tsuei and J.R. Kirtley \cite{tsu00}.
\item{}Stripes:-For recent overview see abstracts
of Stripes 2000 conference. In our way of thinking
stripes arise out of strongly correlated 1-d systems
which are charge transfer insulators, a specific case
of Mott Insulators. Thus as AF Mott Insulating state
is doped, see Fig.~1 it undergoes transition to a strange 
metal[SM], to normal metal [NM] and superconducting
state [SC]. We note that the bandgap in the Mott 
Insulating state is on the order of 2 eV, and
SC $\Delta$ is on the order of meV.
\item{}1/8 problem:-The recent experimental
work of Koike et al.~\cite{koi00} indicates that the
dynamical stripe correlations of holes and spins exist
in Bi-2212, Y-123 and also La-124 and that they tend
to be pinned by a small amount of Zn at 
$p \sim 1/8$\footnote{where p is the hole concentration per Cu}, 
leading to 1/8 anomaly.
\item{}Pseudogap:-We can consider the reduction
of the density of states near the Fermi energy
as a pseudogap. For example Nakano et al.\cite{nak98}, claim
that magnetic susceptibility measurements of
the cuprate La$_{2-x}$ Sr$_{x}$ Cu O$_{4}$ [LSCO]
in the T-x phase diagram show two crossover lines
T$_{max}$(x) and T$^{*}$ (x) [where T$_{c}$ $<$ T$^{*}$
$<$ T$_{max}$], see Fig.~1. Thus these lines T$_{max}$(x) and 
T$^{*}$ (x) are naturally termed the high and low energy pseudogap
respectively. These lines in the T-x\footnote{x is same as
$\delta$} diagram are both montonically decreasing with rising
hole concentration $x$. Below T$_{max}$ magnetic
susceptibility exhibits a broad peak which in the 
usual interpretation is taken to arise from the 
gradual development of the antiferromagnetic spin
correlation.  The lower crossover line T$^{*}$
is taken to represent the temperature below which
a spin gap opens up in the magnetic excitation
spectrum around $q=(\pi,\pi)$. 
\item{}Carrier Inhomogeneity: We have previously emphasized
the carrier [electron] inhomogeneity in our modelling of
HTSC materials \cite{alam99-1} material. In contrast many
models of HTSC assume that charge carrier introduced by doping
distribute uniformly, leading to an electronically homogeneous
system, as in normal metals. However recent experimental
work \cite{pan01} confirms our intuition, which is encouraging.
This inhomogeneity is expected to be manifested in both
the local density of states spectrum and superconducting gap.

\end{itemize} 

\subsection{Luttinger Liquid}
	For the benifit of the non-theoretical reader we now
summarize and state some relevant definitions, formulas and 
properties of Luttinger Liquid [LL], for more details 
see \cite{jon95,jon00}. Moreover, since Single-Walled Nanotube [SWNT]
is involved in our suggested configurations, see section II,
we give some relevant details of SWNT.
\begin{itemize}
\item{}Experimentally, one place where LL behavior has 
been reported, is in the context of carbon nanotubes by
Bockrath et al., \cite{boc99}. LL behavior arises out
of strong electron correlations, thus if $g=(1+2U/\Delta)^{-1/2}$ 
measures the strength of interactions between electrons,
where $U$ is the charging energy of the nanotube and 
$\Delta$ is the energy-level spacing. Very simply
the LL is {\em one-dimensional} correlated electron
state, parameterized by coupling $g$. $g=1$ gives the case
of non-interacting electrons and $g<<1$ represents strong
repulsive interactions.
\item{}The transport properties of LL are dramatically
different, for example, from a Fermi Liquid [FL]. 
In a FL tunnelling amplitude of an electron is energy
independent, in contrast in a clean LL we expect that
 tunnelling amplitude is expected to show a power law
behavior. This implies that conductance $G$ of LL at
small biases [i.e. $eV << k_{_B}$] is dictated by
\begin{equation}
G(T) \propto T^{\alpha}
\label{e1}
\end{equation} 
for large biases [i.e. $eV >> k_{_B}$] the differential
conductance is given by
\begin{equation}
\frac{dI}{dV} \propto V^{\alpha}.
\label{e2}
\end{equation}
Clearly the crucial parameter is $\alpha$, which depends on the
number of the one-dimensional channels, and on where the electron
tunnels to, the bulk or the end of the LL. Thus for a single-walled
nanotube [SWNT], with four conducting modes at the Fermi energy
$E_{_F}$
\begin{eqnarray}
\alpha_{bulk} &=& (g+g^{-1}-2)/8,\nonumber\\
\alpha_{end} &=& (g^{-1}-1)/4.
\label{e3}
\end{eqnarray}
Clearly if $g=1$ we obtain the non-interacting case with
$\alpha$ vanishing. 
\end{itemize}

\section{Suggested Configurations}
\subsection{SET}

\begin{itemize}
\item{}Definition: In the most basic form the definition 
of SET reads \cite{ave00}: SET transistor is a small conductor
[typically a metallic island] placed between two bulk
electrodes, that forms two tunnel junctions with these
electrodes.
\item{}{SET as Quantum detectors} SET is a natural measuring device 
for {\em charge states} mesoscopic Josephson junctions [JJ]: One 
application is in potential quantum logic circuits. We suggest:
To measure Charge fractionalization [CF] in cuprates and
other charge nanodynamics in HTSC and related 
materials.

To realize this we suggest three possible configurations
below. Encouraging is the claim by Averin that in
the advantage of co-tunneling for quantum signal detection is
a {\em weaker} back-action noise on the measured system 
produced by the SET transistor.

\item{}{SET/FET}:A room temperature! SET has been
demonstrated by Mutsomoto\cite{mat00}. Coulomb oscillations
are seen from which the gate capacitance is determined
to be $8\times 10^{-20}$ F. Since we are mainly interested 
in the action of the SET device at lower temperatures the 
width of the gate insulator for sidegate SET can be much 
lower than 964 nm, which is needed at room temperature to 
prevent gate leakage current, for example, it can be on the 
order of 300 nm.
\end{itemize}
\subsection{Configurations}

\begin{itemize}
\item{}SET-JJ:SET coupled to HTSC Josephson junction:-The
SET works on Coulomb energy and on the process of 
tunneling. 
The Coulomb charging of the island takes place due to 
electron tunneling. The current $I$ in this system 
depends on the electrostatic potential of the island 
which in turn is controlled by external gate voltage $V_{g}$.
The SET transistor operation as a detector entails
the measurement of the variations of the voltage  
$V_{g}$ which is sensitive to the current $I$.
It is natural to consider the SET as a quantum 
detector \cite{ave00} since the SET transistor is the
natural measuring device for the potential quantum
logic circuits based on the charge states of 
mesoscopic Josephson junctions \cite{ave00}.
Moreover it is natural to consider the SET as a 
quantum detector if one carefully consider its noise 
properties \cite{ave00}, since it appears that in
the co-tunneling region, for quantum signal detection 
a {\em weaker} back-action noise is produced by the SET 
transistor on the measured system. We propose to couple
the SET transistor to HTSC Josephson junction
and study the charge dynamics of the HTSC
materials for various levels of doping. 
The `insulating' layer in the Josephson 
junction is chosen for a particular value of
doping. For example, in LSCO system it is known
that that superconductivity is suppressed
to some extent at hole concentration per Cu
of 1/8. This is considered to be due to
existence of stripes. One would expect measurable 
change in SET current as one goes from this region
towards the `purely' AF-phase Region I] and `purely' 
superconducting, Fig.~1. 
In principle we can explore all the underdoped
region of T-$\delta$ phase diagram, Fig.~1 by using
different values of doping.

	It is thought that in AF [Region I, Fig.~1] diagonal
stripes develop, whereas in Region II one has vertical
stripes, which allow charge transport. It would be
possible with this configuration to see the evolution
of diagonal to vertical stripes, as doping is varied
and one crosses from the AF to the SM and SG region. 

\item{} SET-SWNT-JJ: This is a slightly more complicated 
version of the previous case. The same remarks apply
here. One of the advantages of this scheme is that we are coupling 
two LL, see Fig.~2. Upto our knowledge this is the
first circuit with LL wires. However, practical problems 
can arise due to the `contacts' which could destroy
the power law dependence a characteristic of LL.
However, it has been recently shown by S.~H\"ugle\cite{hug00}
in the study of electron tunneling from a tip or a lead
into an interacting quantum wire described by LL that
the dynamic image potential is not strong enough to
alter the power-law exponents entering the tunneling
density of states.
\item{}SET transitor coupled to `HTSC material' 
SET transistor:- In this set-up we propose to 
couple a SET transitor to a `HTSC material' 
SET transistor. In the simplest case the
`metallic' island in ordinary SET is replaced
with `HTSC material' for a value of doping
which is in the region of T-$\delta$ phase diagram
which corresponds to metallic phase and `strange' 
metallic [SM] phase. The SM phase is ascribed to 
coexistence of superconductivity and stripe phases. 
In this region the material, if it were perfectly 
oriented would be a superconductor in one direction
and a strongly-correlated insulator in
the other. Thus one can quantify the SM and
metallic phases of HTSC materials in
detail by using charge transport properties
measured with SET. 

	It is known that in the superconducting tunnel 
junctions coherent tunneling of Cooper pairs competes with the
Coulomb blockade \cite{ful89}. Thus one
must consider how the shot noise could be affected 
due to quasiparticle scattering and coherence of the 
supercurrent. Choi et al.\cite{cho01} have considered the 
shot noise in superconducting SET near a resonance Cooper 
pair tunneling. The set-up is typical, as shown in Fig.~\ref{fig3}, where 
two small tunnel junctions are in series, with a small central 
electrode. This system couples capactively to a gate
electrode and via tunneling to the two leads, Fig.~\ref{fig3}.
One intuitively expects that the shot noise [zero-frequency]
must be suppressed if there is coherent oscillation of
the Cooper pairs in the presence of Coulomb blockade.
This is indeed found to be the case \cite{cho01}.
Keeping this in mind we can compare the coherent oscillation
of the Cooper pairs in the SM and full superconducting state.
In the SM case where the superconducting coexists with stripe
phase the coherent oscillation would have a weaker effect
on the shot noise compared to the case where HTSC
superconducting state exists. In this manner by using
the dependence of shot noise on coherent Cooper pair
tunneling we can understand qualitatively and quantitatively
the similarities of superconductivity in SM phase and
in the `full' superconducting phase.  
\item{}SET-Single-Cooper pair box:- Tsai et al. \cite{tsa00}, 
have recently demonstrated the time and energy domain
response of an artificially constructed two-level 
system, which is expected to form one of the
possibilities for the basic bit of quantum computing
[Qubit]. This device which has submicron size allows
one to observe quantum coherent oscillations in a
solid state system whose quantum states involve
a macroscopic number of quantum particles. 
As already mentioned it has already  been noted
\cite{ave00} that the SET is a natural measuring device 
for the potential quantum logic circuits based on the 
charge states of mesoscopic Josephson junctions, 
such as Single-Cooper pair box of Tsai et al.~\cite{tsa00}. 
Keeping in mind that HTSC materials are {\em doped} Mott 
insulators [unlike the low-temperature superconductors], and
consequently their superconductivity depends on
the level of doping as is clear from the T-$\delta$ phase
diagram, Fig.~1, so that one has underdoped, optimally
doped and overdoped regions. It is interesting
to think of an experiment that can give us a detailed
look at the effect of changing T$_{c}$ [as doping
is varied] on the charge transport. A possible
experimental set-up for this measurement could 
consist of SET transistor coupled to a HTSC single-Cooper 
pair box. In this configuration the SET is in essence
coupled to a two-level system. An equivalent circuit
of Cooper-pair box is shown in Fig.~\ref{fig4}, with 
the condition $J_{_{L}}=E_{_{J}}$, $J_{_{R}}=0$, and
one of the lead has to be used as a 
probe gate [i.e. $V_i=2\Delta/e$] the other is set
to zero [i.e. $V_i=0$], $i=L,R$. Here $J_i$ is the
Josephson coupling energy. In this suggested set-up we
can in principle monitor the charge dynamics in  
the whole range of doping when the material is
in the SC state, i.e. Region III, in Fig.~1. We can
study by such an experiment if there is a change in
charge transport as we vary the doping but remain within
region III, in Fig.~1.     
\end{itemize}

\begin{figure}
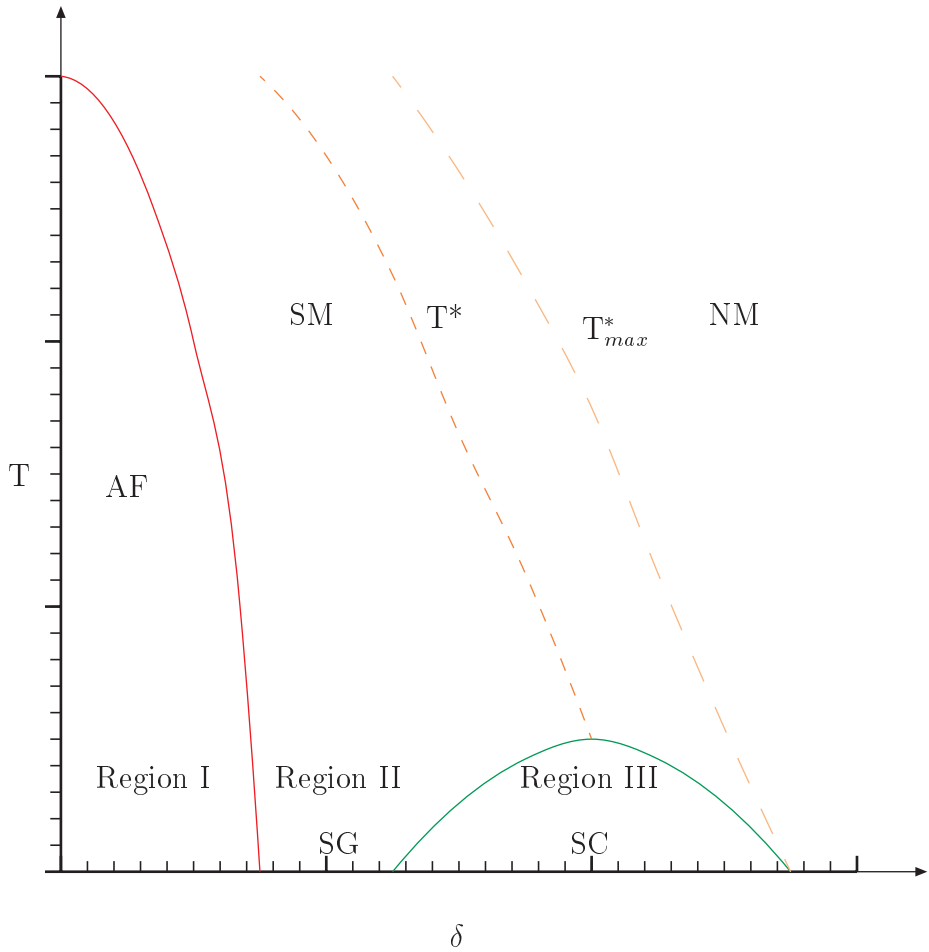

\caption{HTSC phase diagram, showing the varied phases exhibited
in temperature [T]- doping $\delta$ plane. The psuedogap curve
T$^{*}_{\rm max}$ is from tunneling data, and T$^{*}$ is from
NMR and specific heat experiments}\label{fig1}
\end{figure}

	Thus given the above set-ups we can in principle explore
the charge dynamics in every region of the rich phase diagram 
of HTSC materials Fig.~\ref{fig1}. Distinctions must be made
between superconducting, normal and strange metallic SETs.
As is clear from our various set-ups we will not only
encounter superconducting, normal and SM SETs but also
hybrids as we probe different regions of HTSC phase
diagram by varying the doping. Yet another challenging problem
is to study magnetotransport of charged stripes in HTSC materials
in our SET based configurations described above.
To characterize the transport properties in striped phase in 
above materials, the experiments must be calibrated against some 
standard, which clearly shows the one-dimensional transport behaviour 
and also must be closely related to
the cuprate superconductors. One such material is
La$_{1.4-x}$ Nd$_{0.6}$Sr$_{x}$CuO$_{4}$ (x=0.1,0.12,0.15)
which is stripe-ordered non-superconducting relative
of HTSC cuprates. Another standard could be the heavily
underdoped YBCO system, where charged stripes can
be detected in the magnetoresistance \cite{lav99}.

\section{Conclusions}
	In conclusion we have proposed that SET based experiments
can detect charge fractionalization in cuprates and related materials.
To this end we have proposed several experimental configurations
which can achieve this goal. However our suggestion is not
limited to charge fractionalization since SET based experiments
can be used to characterize and understand the underlying physics 
of charge transport and charge fluctuations in these and related
materials for a range of doping. Experiments such as these can be 
classed in a general sense as mesoscopic and nano characterization 
of cuprates and related materials. One of the main point of our
proposal can be summarized thus: If we follow the tunneling of an 
electron, we will `partially' loose the electron as it encounters
a strongly correlated electron state and our SET will register a 
fraction of an electron.

\section*{Acknowledgments}
The Sher Alam's work is supported by the Japan Society for
for Technology [JST].

\begin{figure}
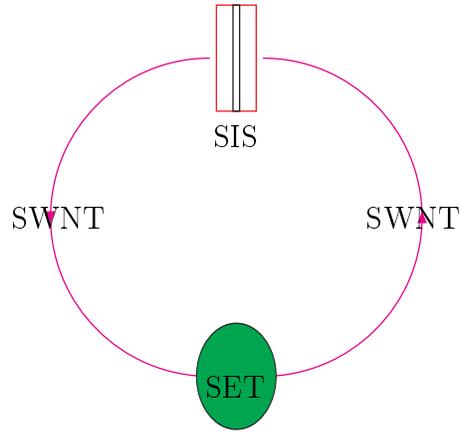

\caption{SET, SWNT, and SIS configuration to measure CF}\label{fig2}
\end{figure}
\begin{figure}
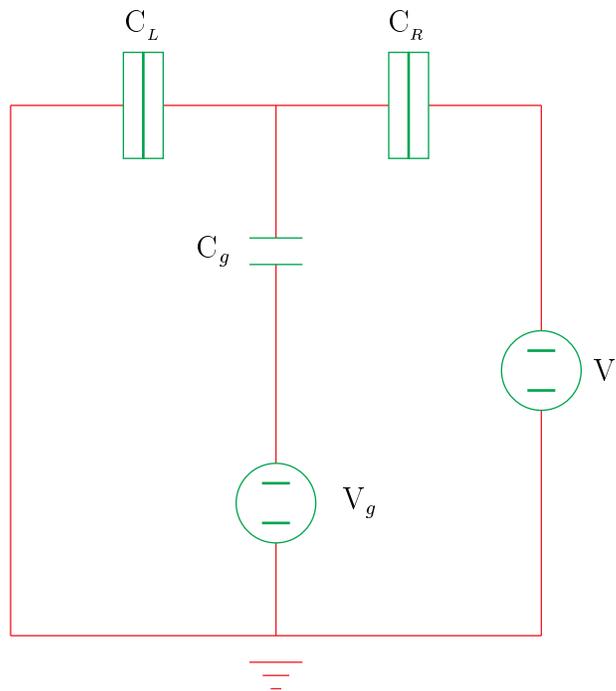

\caption{Schematic diagram of Superconducting SET}\label{fig3}
\end{figure}
\begin{figure}
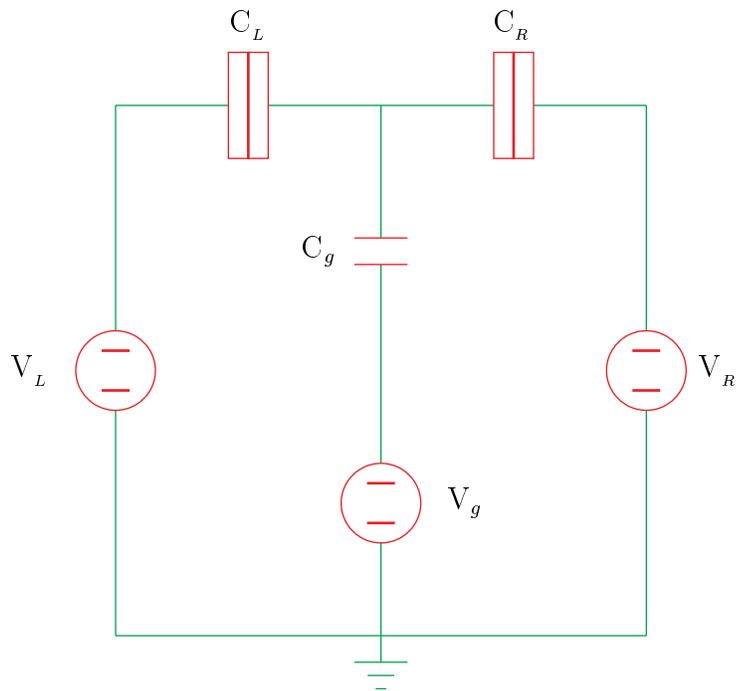

\caption{Equivalent circuit of Cooper-pair box}\label{fig4}
\end{figure}
\end{document}